\documentclass{PoS}
\PoS{PoS(LAT2005)309}

\usepackage[centertags]{amsmath}
\usepackage{amsfonts} 
\usepackage{amssymb} 
\usepackage{amsthm}
\usepackage{graphicx}
\usepackage{psfrag}
\usepackage{cite}
\def\one{{\rm 1\kern -.9mm l}}

\def\beq{\begin{equation}}
\def\eeq{\end{equation}}
\def\beq{\begin{equation}}
\def\eeq{\end{equation}}
\def\beqa{\begin{eqnarray}}
\def\eeqa{\end{eqnarray}}
\newcommand{\bra}{\langle}
\newcommand{\ket}{\rangle}
\newcommand{\eq}[1]{eq. (\ref{#1})}

\newcommand{\Z}{\mathbb{Z}}

\title{QCD string from D0 branes }

\ShortTitle{QCD string from D0 branes }

\author{Marco Bill\'o,~~~\speaker{Michele Caselle}\\
Dipartimento di Fisica Teorica, Universit\`a di Torino\\
and Istituto Nazionale di Fisica Nucleare - sezione di Torino,\\
Via P. Giuria 1, I-10125 Torino, Italy\\
        E-mail: \email{billo,caselle@to.infn.it}}

\author{Martin Hasenbusch\\Dipartimento di Fisica dell'Universit\`a di Pisa
 and I.N.F.N.,\\
 Largo Bruno Pontecorvo 3, 
                  I-56127 Pisa,
                        Italy  \\
        E-mail: \email{Martin.Hasenbusch@df.unipi.it}}

\author{Marco Panero\\School of Theoretical Physics,
Dublin Institute for Advanced Studies,\\
 10 Burlington Road, Dublin 4,
                              Ireland\\
        E-mail: \email{panero@stp.dias.ie}}

\FullConference{XXIIIrd International Symposium on Lattice Field Theory\\
 25-30 July 2005\\
 Trinity College, Dublin, Ireland}
\abstract{We report the results of a set of high precision simulations performed
 in the 3d gauge Ising model. We evaluated  the interquark potential and the
first few energy levels and compared them with the predictions obtained with the effective Nambu-Goto string and
with the free bosonic string. The data are precise enough
to unambiguously 
distinguish between the free string predictions and those obtained using the N-G
effective string. At large distances we find a remarkable agreement between 
Monte Carlo data and N-G predictions for the first excited 
energy level, while the free string picture is definitely excluded.
As the interquark distance is decreased (and/or the finite temperature becomes
higher) the Monte Carlo results show larger and larger deviations both from the N-G and from the free
string predictions. In order to better understand this behaviour we
re-derived the effective Nambu-Goto theory result for
the Polyakov loop correlator using a
covariant quantization. 
We chose as boundary conditions those of an open string
attached to two D0-branes at spatial distance $R$, in a target space with
compact euclidean time. Obviously our treatment is fully consistent only in
$d=26$. The extension to generic $d$ requires taking into account the Liouville mode of
Polyakov's formulation. The analogy with the standard light cone calculation suggests that
the contribution due to the Liouville field can be neglected for large $R$. 
At shorter scales, the Liouville mode cannot be neglected and 
its contribution to the interquark potential might be the source of the discrepancies with respect to the effective N-G results that we observe in our Monte Carlo simulations.}

\begin{document}

\section{Introduction}
\label{sec:intro}

The starting point of the effective string description of the 
interquark potential is to model the latter in terms of a string
partition function. Choosing for instance to evaluate the potential by using 
the expectation value of a pair of Polyakov loops we have:

\beq
\label{prp0conazeff}
G(R,L) = \bra P^\dagger (R) P(0) \ket = \int \left[ \mathcal{D} h \right] e^{-S_{\mbox{\tiny{eff}}}} \equiv Z(R,L)~,
\eeq
where $L$ is the length of the lattice in the compactified direction (i.e. the inverse temperature $L\equiv 
1/T$) and
$S_{\mbox{\tiny{eff}}}$ denotes the effective action for the world sheet spanned by the string. 
In eq.~(\ref{prp0conazeff}), the functional integration is done over world sheet configurations 
which have fixed boundary conditions along the space-like direction, and periodic boundary conditions 
along the compactified, time-like direction (the Polyakov lines are the fixed boundary of the string 
world sheet).

The simplest and most natural string model is the Nambu-Goto 
one, which assumes the string action $S_{\mbox{\tiny{eff}}}$ to be proportional to the area spanned 
by the string world sheet:
\beq
\label{azionenambugoto}
S_{\mbox{\tiny{eff}}}= \sigma \cdot \int d^2 \xi \sqrt{ \det g_{\alpha \beta} }~.
\eeq 

For the Nambu-Goto string with boundary conditions corresponding to fixed
ends in the spatial directions (the static quark and anti-quark) Alvarez~\cite{alvarez81} (for $d\to\infty$) and Arvis~\cite{Arvis:1983fp}, with a formal
quantization, obtained the energy spectrum 
\beq
E_n(R)=\sigma R \sqrt{1+\frac{2\pi}{\sigma R^2}\left(n-\frac{d-2}{24}\right)}~.
\label{en}
\eeq
 The partition function can thus be written as
\beq
\label{zng}
Z(R,L)=\sum_n w_n e^{-L E_n(R)}~,
\eeq
$w_n$ being the usual multiplicities of the  bosonic string. 
The static potential coincides with the  lowest energy level: $V(R) = E_0(R)$,
and reproduces the well-known ``L\"uscher term''~\cite{lsw,l81} in the large $R$ limit.
Indeed in this limit the Nambu-Goto effective action becomes the 2d CFT of
$(d-2)$ massless, non-interacting bosons. 

The derivation of
eq.s~(\ref{en}) and~(\ref{zng}) in~\cite{Arvis:1983fp} uses the
re-parametrization invariance of the world-sheet to reach the conformal gauge
(where the Nambu-Goto action is equivalent to the free string action) and the
residual conformal invariance to fix a light-cone type gauge (which is sometimes
denoted as ``physical gauge''). This leaves the transverse modes as the only independent dynamical variables, which become oscillators upon quantization. 
This bosonic string model, of course, is truly consistent at the quantum level
only if $d=26$: as usual in light-cone type gauges, Lorentz invariance is
otherwise broken. It was however noticed in~\cite{Olesen:1985pv} that the
coefficient of the anomaly vanishes for $R\to\infty$, so that in this regime the
model could be consistent. 

In these last years, thanks to  various remarkable improvements in lattice 
simulations,
the effective string picture could be tested with a 
very high degree of precision and confidence~\cite{cfghp97,chp03,lw02,lw04,chp04,Panero:2005iu,cpr04,chp05,chpnew,m02,jkm03,jkm04} supporting the presence of a L\"uscher-type correction  
in the interquark potential at large enough inter-quark
distances (and low temperatures). Moreover, this result seems to be characterized by a high degree of universality, 
meaning that it does not depend on the particular gauge group under
study (the same behaviour is observed in models as different as the
$\mathbb{Z}_2$ gauge model in $(2+1)$ dimensions~\cite{chp05} and the
$\mathrm{SU}(3)$ LGT in $(3+1)$ dimensions~\cite{lw02}).
However this result in itself is not very informative since it simply indicates
that the quantum fluctuations of the flux tube behave as free bosonic degrees of freedom at large distance, but
gives no insight on the nature of the effective string which describes the flux tube. To this end it would be
important to be able to study the higher order corrections due to the effective string. A task which is very 
difficult to pursue as far as the interquark potential is concerned but becomes more feasible if one looks at the
excited energy levels.
The aim of this contribution is to address this point following two lines:
\begin{itemize}
\item
First we shall discuss some recent numerical tests of the energy spectrum 
of the string performed in the 3d gauge Ising model. 
The picture which emerges from these simulations
is that the Nambu-Goto effective string prediction \eq{en} correctly
describes the Monte Carlo data at large enough distance, 
while the free string model alone is incompatible with the numerical results.
At the same time it is clear from the data that as the
inter-quark distance decreases and/or the temperature increases (i.e. as the
de-confinement transition is approached) large deviations from the N-G predictions appear
(and also the universality mentioned above is partially
lost)~\cite{chp04,chp05,jkm03,jkm04}.

\item
Second,  we shall describe an alternative procedure to derive the
effective Nambu-Goto theory,  starting from the free
bosonic string and using a covariant quantization~\cite{bc05}. 
This derivation makes more transparent 
the limits involved in the construction of the effective
action starting from the string lagrangian and may help to understand the observed deviations at short distance.
\end{itemize}

\section{Numerical results}
\label{sec:NG}

In order to test the ability of the  Nambu-Goto effective string to describe the interquark potential we
concentrated on the $\Z_2$ pure lattice gauge theory in $d=3$, which is expected to provide a 
prototypical model for quark confinement. This choice is also supported by the observation that, in the region
where the two models can be compared,  the Ising and the SU(2) LGT's in (2+1) dimensions
show very similar behaviours both from a qualitative and a quantitative point of view~\cite{cpr04}.
The nice feature of this choice is that by using duality one is able to
obtain very precise estimates for the ratios of two subsequent Polyakov loop correlators: $G(R+1,L)/G(R,L)$.
By integrating these ratios from $R=0$ to $R=R_{\mbox{\tiny{max}}}-1$ we may then directly obtain the correlator 
$G(R_{\mbox{\tiny{max}}},L)$ and hence the effective string partition function $Z(R_{\mbox{\tiny{max}}},L)$.

\begin{figure}
\centerline{\includegraphics[height=100mm]{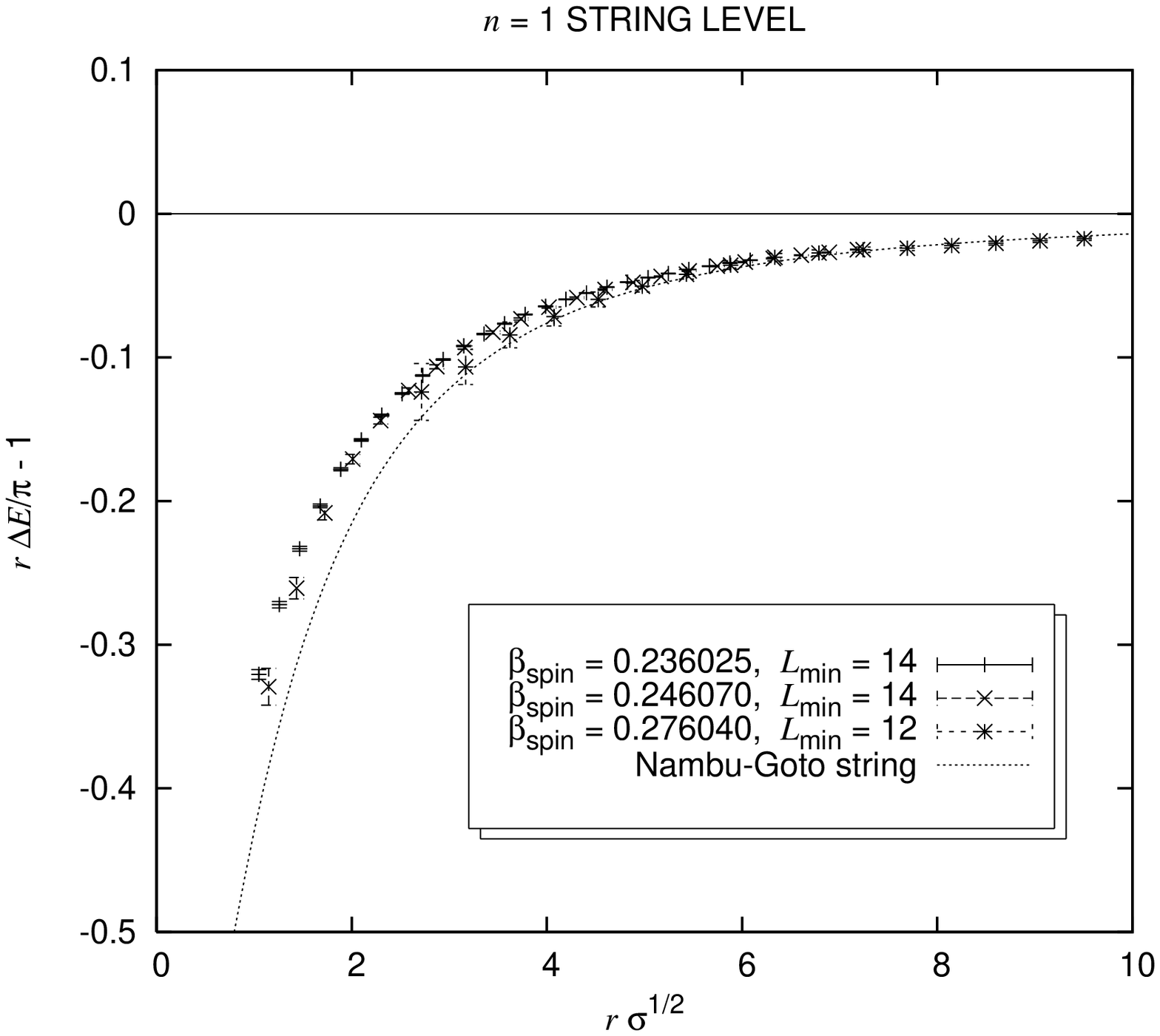}}
\vspace{1cm}
\caption{Relative deviation of the first energy gap $E_1 - E_0$, with respect to the free string prediction, 
as a function of the interquark distance; results are shown for different values of $\beta_{spin}$, and are 
scaled in physical units.}
\label{r04redchisqvslminfig}
\end{figure}

Our results are reported in figures~\ref{fig:ising} and~\ref{r04redchisqvslminfig}.
The first plot shows the deviation of ratios of Polyakov loop correlators from the free string model, as a function of $L$. Looking at the figure, 
one can appreciate the agreement with N-G at large distance and the 
deviations as $L$ decreases. Thanks to the peculiar geometry of the observable (since $L\equiv 1/T$
as $L$ decreases we move toward the deconfinement transition where the higher order terms in the string action
play a more and more important role) 
one can also appreciate the difference between the N-G predictions and the free string ones. 

As anticipated above, this difference is better appreciated in fig.~\ref{r04redchisqvslminfig}
in which we plotted the relative deviation of the first energy gap $E_1 - E_0$ with respect to the free 
string prediction, as a function of the interquark distance. The data disagree from the N-G
expectation for low values of $R$ and then nicely converge toward it as $R$ increases. It is interesting to
observe that our data perfectly agree, for low values
of $R$,  with those reported in~\cite{jkm04} where in fact a disagreement with respect to 
the N-G picture was claimed. We confirm this disagreement at short distance, but being able to extend our
analysis to larger values of $R$ we can confirm (in agreement with the above 
observations) that at large distances the N-G picture is fully restored. The solid, horizontal line in 
fig.~\ref{r04redchisqvslminfig} represents the free string prediction, which is 
clearly excluded for all values of $R$.

More details can be found in the original papers~\cite{chp04,chp05,bc05,chpnew}.

\FIGURE{
\includegraphics[width=12cm]{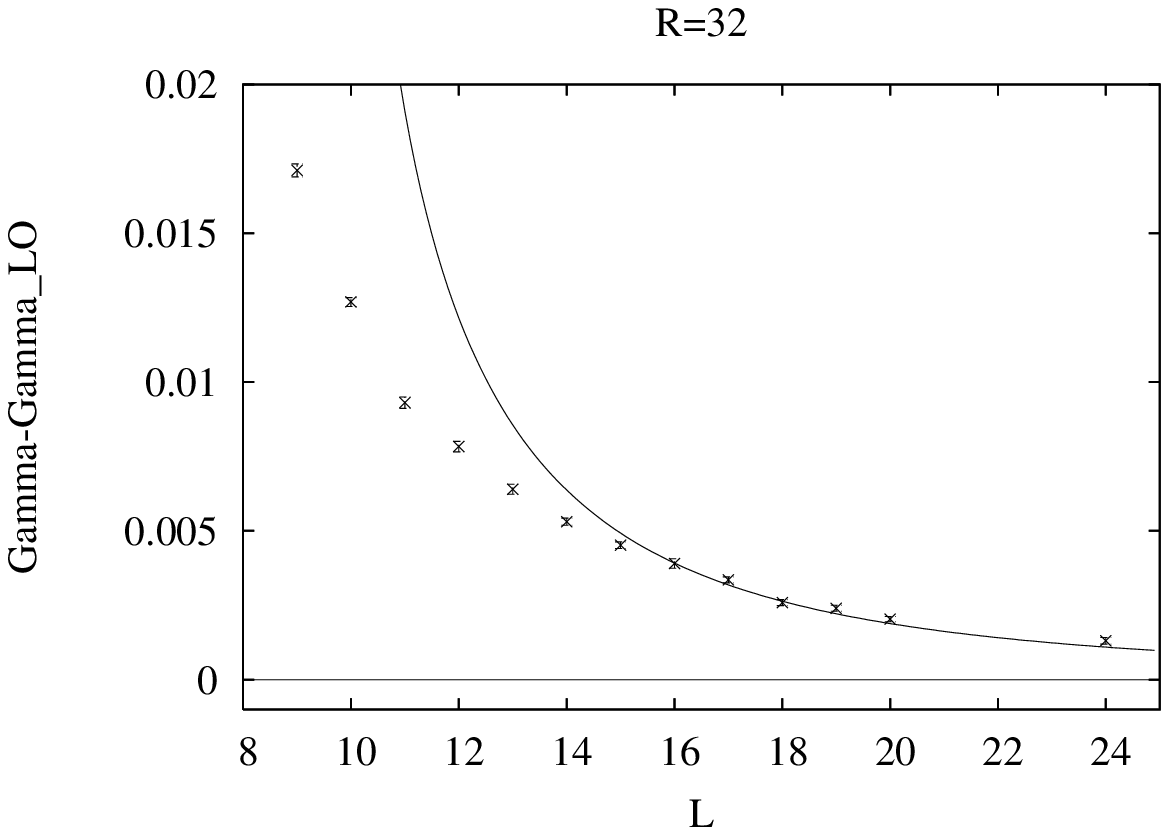}
\caption{\small Monte Carlo results for the Polyakov loop correlators in the
(2+1) dimensional gauge Ising model ($\beta_{spin}=0.226102$). The data are taken at a fixed value  $R=32$
of the interquark distance and a varying size ($8<L \leq 24$) of the lattice in the
time direction. The figure shows the deviation of $\Gamma$ (the ratio
$G(R+1)/G(R)$ of two correlators shifted by one lattice spacing, see~\cite{chp05} for details)
with respect to the asymptotic free string expectation
$\Gamma_{LO}$ (which with this definition of observables corresponds to the
straight line at zero). The curve is the Nambu-Goto prediction for this
observable. Notice the remarkable agreement in the range $16 \leq L \leq 24$, which 
is not the result of a fitting procedure: in the
comparison reported in the figure there is no free parameter (taken from fig.~3
of ref.~\cite{chp05}).}
\label{fig:ising}
}

\section{Effective string from D0 branes}
\label{sec:d0}
The Nambu-Goto effective string spectrum of eq.~(\ref{en})
can also be obtained following an alternative route,
using the reparametrization invariance to reach the conformal gauge 
(where the N-G action is equivalent to the free
string action) and then quantizing the model by using the so called 
covariant quantization.  
The boundary conditions of the string are described in modern
terms as those of an open string attached to two D0-branes (which play the role of the Polyakov loops)
at spatial distance $R$. In this framework, the anomaly 
shows up with
the appearance of  an
additional field (the so called ``Liouville mode''). If one assumes that at large distance this mode
can be neglected, then one can quantize the string as it is usually 
done for the critical bosonic string, and re-obtain all the 
previous results (see~\cite{bc05}). In particular, one finds exactly the 
Nambu-Goto spectrum of eq.~(\ref{en}). The major outcome of 
this alternative procedure is that it makes 
the role of the Liouville mode explicit, and 
it could in principle offer a clue to guess its contribution to the effective string
prediction
at shorter interquark distances and possibly explain the deviations observed in the lattice simulations in this
regime. Besides this
we think that this alternative derivation of the Nambu-Goto effective theory 
 may have some further important advantages. 
 Not having fixed a
``physical gauge'', the world-sheet duality between the open and closed channel
is most evident and allows for an explicit interpretation of the free energy in
terms of tree level exchange of closed string states between boundary states.
Moreover our formulation is well suited in principle to study the contributions to the
inter-quark potential from string interactions, which in our language would mean
wrapping the Polyakov string on surfaces (bordered by the Polyakov loops) with
handles.  Finally,  it opens the possibility of investigating the  relevance
in the gauge theory of the contributions to the free energy with different
winding numbers (which appear naturally in our framework).

\end{document}